\documentclass[acmtog]{acmart}
\acmSubmissionID{579}

\usepackage{booktabs} 
\usepackage{graphicx}
\usepackage{subcaption}
\usepackage{caption}

\usepackage[ruled]{algorithm2e} 

\SetAlFnt{\small}
\SetAlCapFnt{\small}
\SetAlCapNameFnt{\small}
\SetAlCapHSkip{0pt}

\copyrightyear{2025}
\acmYear{2025}
\setcopyright{rightsretained}\acmConference[SIGGRAPH Conference Papers '25]{Special Interest Group on Computer Graphics and Interactive Techniques Conference Conference Papers }{August 10--14, 2025}{Vancouver, BC, Canada}
\acmBooktitle{Special Interest Group on Computer Graphics and Interactive Techniques Conference Conference Papers (SIGGRAPH Conference Papers '25), August 10--14, 2025, Vancouver, BC, Canada}
\acmDOI{10.1145/3721238.3730665}
\acmISBN{979-8-4007-1540-2/2025/08}

\begin{document}
\title{Motion Control via Metric-Aligning Motion Matching}

\author{Naoki Agata}
\orcid{0009-0007-6443-4907}
\affiliation{%
  \institution{The University of Tokyo}
  \city{Bunkyo-ku, Tokyo}
  \country{Japan}
}
\email{agata-naoki@is.s.u-tokyo.ac.jp}

\author{Takeo Igarashi}
\orcid{0000-0002-5495-6441}
\affiliation{%
  \institution{The University of Tokyo}
  \city{Bunkyo-ku, Tokyo}
  \country{Japan}
}
\email{takeo@acm.org}

\begin{abstract}
We introduce a novel method for controlling a motion sequence using an arbitrary temporal control sequence using temporal alignment. Temporal alignment of motion has gained significant attention owing to its applications in motion control and retargeting. Traditional methods rely on either learned or hand-craft cross-domain mappings between frames in the original and control domains, which often require large, paired, or annotated datasets and time-consuming training. Our approach, named \textit{Metric-Aligning Motion Matching}, achieves alignment by solely considering within-domain distances. It computes distances among patches in each domain and seeks a matching that optimally aligns the two within-domain distances. This framework allows for the alignment of a motion sequence to various types of control sequences, including sketches, labels, audio, and another motion sequence, all without the need for manually defined mappings or training with annotated data. We demonstrate the effectiveness of our approach through applications in efficient motion control, showcasing its potential in practical scenarios.
\end{abstract}

%
%

\ccsdesc[500]{Computing methodologies~Motion processing}

%
%

\keywords{character animation, motion control, motion editing interface, optimal transport}

\begin{teaserfigure}
    \includegraphics[width=\textwidth]{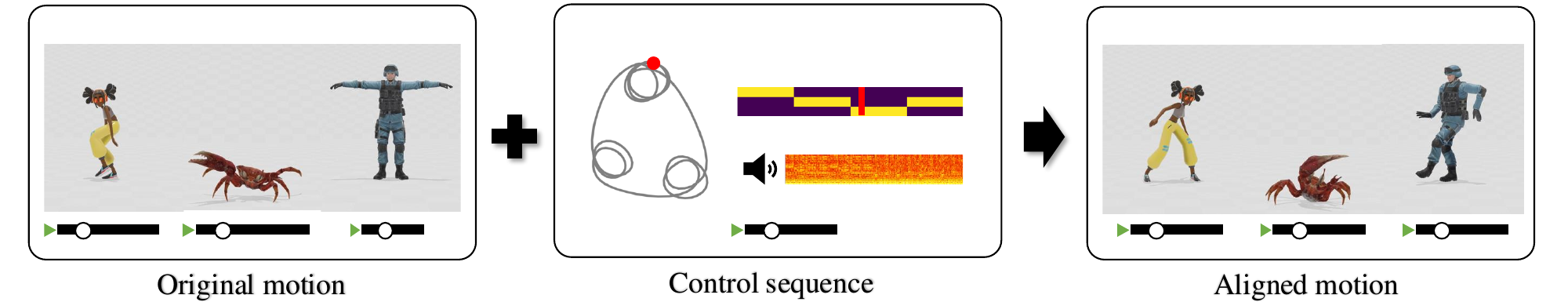}
    \caption{Overview of our method. Our method aligns given original motion to given control sequence.
    Our method can take arbitrary control sequences, such as sketches, labels, audio, and motions. 
    Our method solely relies on within-domain distance in original and control, without needing manual 
    definition of mapping or training with annotated data.}
    \Description{This is teaser figure for the article.
}
    \Description{This is the teaser figure for the article.}
    \label{fig:teaser}
\end{teaserfigure}

\maketitle

\section{Introduction}
\label{sec:intro}

The problem of aligning a motion sequence with a control sequence has gained increasing attention owing to the wide range of tasks it enables. Applications include, for instance, controlling motion based on a user’s dynamic input, retargeting human movement to a character with a different skeletal structure, and synchronizing dance motion with audio structure. However, these tasks present significant difficulties, often demanding precise temporal correspondence and structural coherence between control sequences and motion sequences, even when the control and motion domains differ significantly in structure, dimension, duration, or complexity.


Early studies on motion-to-motion retargeting relied on hand-crafted mapping functions and strategies~\cite{gleicher_retargetting_1998,lee_hierarchical_1999,choi_online_2000,tak_physically-based_2005,yamane_animating_2010,rhodin_generalizing_2015,seol_creature_2013}. More recent approaches leverage large-scale learning-based frameworks to learn direct mappings between two motions~\cite{villegas_neural_2018,lim_pmnet_2019,li_ace_2023} or shared embeddings between two domains~\cite{aberman_skeleton-aware_2020, lee_same_2023, li_walkthedog_2024}. While deep learning techniques compute mappings empirically, they often require extensive annotated datasets and prolonged training to ensure robust generalization. Similarly, audio-driven motion synthesis~\cite{alexanderson_listen_2023,tseng_edge_2023,li_lodge_2024} depends heavily on large paired datasets and substantial computational resources. Recent efforts to reduce data demands, as seen in works like GANimator~\cite{li_ganimator_2022}, GenMM~\cite{li_example-based_2023}, and SinDDM~\cite{raab_single_2023}, demonstrate that motions can be generated from limited data. However, these methods still lack robust, built-in alignment strategies for complex temporal inputs and often require domain-specific constraints or supplementary annotations to measure cross-domain correspondences.

In this study, we propose a motion alignment technique that relies solely on within-domain distances, which can be naturally defined without requiring hand-crafted or learned mapping functions (Fig.~\ref{fig:teaser}). Our approach first computes distances among patches within the original motion and control sequences separately. We then search for couplings between patches in the control sequence and those in the original motion such that the within-domain distance structure is preserved as much as possible. These couplings are used to compute a weighted blend of original patches, resulting in an aligned motion. Technically, our method uses metric alignment techniques based on the fused semi-unbalanced Gromov-Wasserstein (FSUGW)  optimization objective~\cite{xu_temporally_2024}, a modified version of the Gromov-Wasserstein optimal transport (GW OT) problem. GW OT has been successfully applied in various applications, such as identifying correspondences between shapes~\cite{solomon_entropic_2016}. By leveraging advancements in GW OT algorithms, our method achieves cross-domain coupling without explicitly computing cross-domain distances. 

This framework enables the alignment of motion sequences to various types of control sequences, facilitating a wide range of applications. These include: (1) waveform-to-motion and sketch-to-motion, allowing users to control motion sequences through parameterized 1D waveform signals and 2D hand-drawn sketches, and (2) motion-by-numbers, where users specify only the temporal segmentation labels. Our framework is also applicable to existing tasks, such as: (3) motion-to-motion alignment, and (4) audio-conditioned motion control.

Our contributions are summarized as follows:
\begin{enumerate}
    \item We introduce \textit{Metric-Aligning Motion Matching (MAMM)}, a unified FSUGW-based optimization framework that aligns motions with a diverse range of control sequences without requiring task-specific training or algorithm redesign.
    \item We demonstrate that our method achieves high-quality alignments between a given motion and control sequence within seconds, eliminating the need for large datasets, extensive annotations, or prolonged training times.
    \item We present novel motion editing techniques---waveform-to-motion, sketch-to-motion, and motion-by-numbers---enabled by MAMM, which, to our knowledge, are the first of their kind.
    \item We demonstrate the generalizability of the MAMM framework to established tasks, such as motion-to-motion and audio-to-motion alignment.  
\end{enumerate}

\section{Related Works}
\label{sec:relwork}
\subsection{Motion Editing}
\label{subsec:rel_motion_editing}
Motion editing techniques aim to modify existing motion data or create new motion that satisfies new spatiotemporal constraints, adheres to specific stylistic requirements, or adapts to different environmental contexts. Early foundational approaches introduced the concept of \emph{spacetime constraints}, formulating motion editing as an optimization problem guided by physical laws~\cite{witkin_spacetime_1988, gleicher_motion_1997}. This line of research enabled animators to enforce physically plausible edits on motion sequences while maintaining user control over keyframed poses.

Subsequent works focused on lowering the barrier for nonexperts and improving workflow efficiency through sketch-based interfaces and other intuitive user controls~\cite{thorne_motion_2004, terra_performance_2004, guay_space-time_2015, choi_sketchimo_2016, garcia_spatial_2019}. These methods allowed users to apply edits by sketching curves or gestures, which were then mapped to character poses and motion adjustments. Another approach involved creating poses by interpolating keyframes spatially~\cite{igarashi_spatial_2005}. Parallel efforts explored physically guided animation tools, where motion was fine-tuned by automatically inferring or imposing physical properties, resulting in more believable outputs~\cite{shapiro_practical_2011, hahn_rig-space_2012}.

More recent solutions integrate optimization frameworks directly into keyframe-based workflows, enabling artists to selectively refine motion segments without fully discarding manual control~\cite{koyama_optimo_2018, ciccone_tangent-space_2019}. In addition, tangible or actuated puppets provide haptic feedback and a physical interface, further bridging digital and real-world motion editing~\cite{jacobson_tangible_2014, yoshizaki_actuated_2011}. User performance itself has also served as an intuitive control mechanism~\cite{dontcheva_layered_2003, lockwood_finger_2012}.

Compared to existing methods, a notable feature of our method is automatically computing the control-to-motion mapping from input sequences in a unified manner. This allows various input methods for motion control. By contrast, existing methods require developers to explicitly define the mapping from control signals to motion.
\subsection{Motion Retargeting and Alignment}
\label{subsec:rel_retarget}
Motion retargeting transfers motion data from a source character to a target character, even when their skeletal structures or proportions differ. Early methods formulated this as an optimization problem, incorporating constraints such as low-level motion representations~\cite{gleicher_retargetting_1998}, kinematic constraints~\cite{lee_hierarchical_1999}, end-effector trajectories~\cite{choi_online_2000}, or dynamic feasibility~\cite{tak_physically-based_2005}. However, these approaches struggled with significant skeletal differences. Later approaches introduced partial correspondences~\cite{yamane_animating_2010, rhodin_generalizing_2015, seol_creature_2013} or learned mappings independent of skeletal structures~\cite{rhodin_interactive_2014} to handle differing topologies.

The advent of deep learning and large motion capture datasets led to data-driven methods~\cite{villegas_neural_2018, lim_pmnet_2019, aberman_skeleton-aware_2020, li_ace_2023, li_walkthedog_2024}. Early learning-based techniques leveraged cycle-consistency and adversarial losses~\cite{villegas_neural_2018} or skeleton-aware components~\cite{aberman_skeleton-aware_2020}. ACE~\cite{li_ace_2023} advanced retargeting for characters with significant skeletal differences using pretrained motion priors. Learning correspondences between general dynamical systems has also been investigated~\cite{kim_learning_2020}.

WalkTheDog~\cite{li_walkthedog_2024} introduced motion alignment via a learned discretized phase manifold, enabling shared embeddings for characters with different topologies, such as humans and dogs. Their method uses frequency-scaled motion matching but requires extensive training and large datasets.

Our method simplifies this process, offering efficient motion alignment for small data pairs without requiring extensive training or large datasets, offering a practical solution for limited-data scenarios.
\subsection{Audio-Driven Motion Synthesis}
\label{subsec:rel_dance}
Audio-driven motion generation (e.g., dance, gesture synthesis) generates character motion from acoustic features, a long-standing topic in computer animation. For a comprehensive review, see~\cite{zhu_human_2024,nyatsanga_comprehensive_2023}.

Early studies on dance synthesis primarily focused on synchronizing dance motions with musical beats, rhythms, and intensity using hand-crafted mappings~\cite{shiratori_dancing--music_2006, shiratori_synthesis_2008, ofli_audio-driven_2008} or employing example-based approaches and statistical models~\cite{ofli_learn2dance_2012, fan_example-based_2012, fukayama_2015_music}. Recent advances in deep learning have enabled the generation of more expressive dance motions by leveraging sophisticated motion generation models~\cite{li_ai_2021, lee_dancing_2019, siyao_bailando_2022, bhattacharya_danceanyway_2024}, including diffusion models~\cite{alexanderson_listen_2023, tseng_edge_2023, li_lodge_2024}. Other approaches combine learned choreo-musical embeddings with motion graphs~\cite{au_choreograph_2022, chen_choreomaster_2021}.

Parallel efforts in speech-driven gesture synthesis have similarly aimed to produce natural co-speech motion. Earlier approaches relied on extensive manual rules~\cite{cassell_beat_2001, kopp_max_2003, lee_nonverbal_2006, lhommet_cerebella_2015}, while more recent methods use learning-based methods~\cite{ferstl_investigating_2018, hasegawa_evaluation_2018, takeuchi_speech--gesture_2017, alexanderson_style-controllable_2020, yazdian_gesture2vec_2021, wu_probabilistic_2021}. However, many of these methods depend heavily on large paired audio-motion datasets~\cite{alexanderson_listen_2023, li_ai_2021, chen_choreomaster_2021, lee_dancing_2019, siyao_bailando_2022, bhattacharya_danceanyway_2024, fan_example-based_2012} and are computationally demanding~\cite{tseng_edge_2023, li_lodge_2024, alexanderson_listen_2023}, limiting their adaptability to new characters or resource-limited environments. Our approach bypasses these challenges by employing fused semi-unbalanced Gromov-Wasserstein optimal transport, which reduces reliance on large datasets and hand-crafted features. This enables more efficient and flexible audio-to-motion synthesis suitable for a wide range of contexts.
\subsection{Motion Synthesis from Small Data}
Building on recent advances in image generation from a single or a few examples~\cite{shaham_singan_2019, granot_drop_2022, kulikov_sinddm_2023}, several studies have explored synthesizing motion using only a limited number of example sequences. For instance, GANimator~\cite{li_ganimator_2022} employs generative adversarial networks to generate diverse motion patterns from a single instance. GenMM~\cite{li_example-based_2023}, in contrast, bypasses the training phase entirely, enabling rapid motion synthesis through a bidirectional similarity-based matching framework. SinDDM~\cite{raab_single_2023} further improves the quality of generated motion using diffusion models.

While these approaches demonstrate significant flexibility in conditioned motion synthesis tasks -- such as inbetweening, trajectory control~\cite{li_ganimator_2022, li_example-based_2023}, and style transfer within the same skeletal structure~\cite{raab_single_2023} -- they often lack inherent mechanisms to handle more complex motion control by sequences. Consequently, achieving desirable outcomes in such scenarios frequently requires additional annotations or manual interventions, which can increase both labor and computational costs.
\subsection{Optimal Transport}
Optimal transport (OT) has been widely applied across various domains, including machine learning, computer vision, and neuroscience~\cite{Montesuma_recent_2024}. For instance, Elnekave and Weiss~\cite{elnekave_generating_2022} introduced methods for generating natural images by matching patch distributions. Advancements in Gromov-Wasserstein (GW) distance have enabled mappings between disparate domains, such as shape correspondence and image-to-color distribution matching~\cite{solomon_entropic_2016}. Recent developments in Unbalanced and Fused GW~\cite{sejourne_unbalanced_2021, chizat_scaling_2018} have further expanded OT applications to areas such as aligning individual brains~\cite{thual_aligning_2022} and unsupervised action segmentation~\cite{xu_temporally_2024}. We incorporate the fused unbalanced GW OT problem solver~\cite{xu_temporally_2024} as a core component of our framework, offering a unified and effective solution to sequence-conditioned motion control.
\section{Method}
\label{sec:method}

\subsection{Motion Representation}

We represent a motion sequence as a series of time-indexed poses, as in a previous study~\cite{li_example-based_2023}, where each pose is described by \textit{root displacement} $V$ and \textit{joint rotation} $R$ in character space. At time $t$, the pose is given by $[V_t, R_t]$, where $V \in \mathbb{R}^{T \times 3}$ represents the changes in the root joint position compared to consecutive frames, and $R \in \mathbb{R}^{T \times J \times 3 \times 3}$ describes the rotation of each joint in the character's root space in matrix representation. Here, $J$ denotes the number of joints.


\subsection{Metric-Aligning Motion Matching}

Our framework, named \textit{Metric-Aligning Motion Matching (MAMM)}, integrates \textit{motion patch extraction and blending}, \textit{FSUGW optimization}, which involves the minimization of \textit{Gromov-Wasserstein (GW) loss} and \textit{Wasserstein loss}, into a unified scheme.

We denote the original motion sequence as $X$ and the control (input) sequence as $Y$, with $X'$ representing the aligned motion sequence (output). The aligned sequence $X'$ is optimized to retain similarity to the original sequence $X$ while aligning with the control sequence $Y$. Both $X'$ and $X$ reside in the same domain, whereas $X'$ and $Y$ often belong to different domains, making it challenging to directly measure the distance between them (e.g., motion and music). Additionally, $X$ and $Y$ may have different lengths. Ultimately, $X'$ and $Y$ are optimized to have the same length and exhibit structural similarity.

\begin{figure}[t]
    \centering
    \includegraphics[width=0.9\linewidth, trim=0 5mm 0 0, clip]{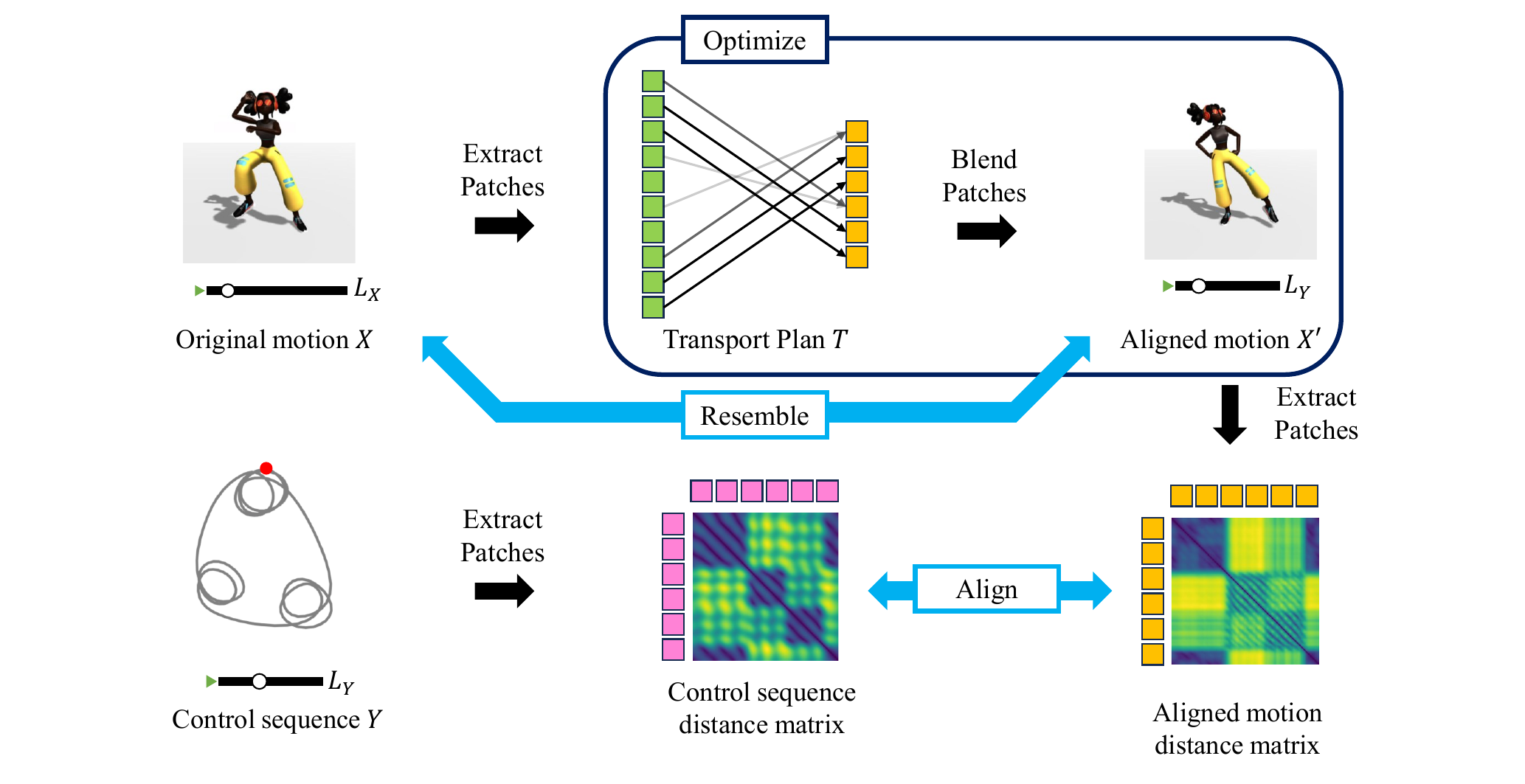}
    \caption{Intuitive concept of MAMM framework. Our framework optimizes transport plan $T$ and aligned motion sequence $X'$, whose patch pairwise distance matrix is similar to that of control sequence $Y$, while resembling original motion $X$.}
    \label{fig:mamm-concept}
\end{figure}

Conceptually, the objective of the MAMM framework is to output an aligned motion such that the pairwise distance matrix of its frames closely resembles the distance matrix of the control sequence while retaining similarity to the original motion (see Fig.~\ref{fig:mamm-concept}). As shown in the figure, the distance matrix of our aligned motion resembles that of the control sequence.

\begin{figure}[t]
    \centering
    \includegraphics[width=0.9\linewidth]{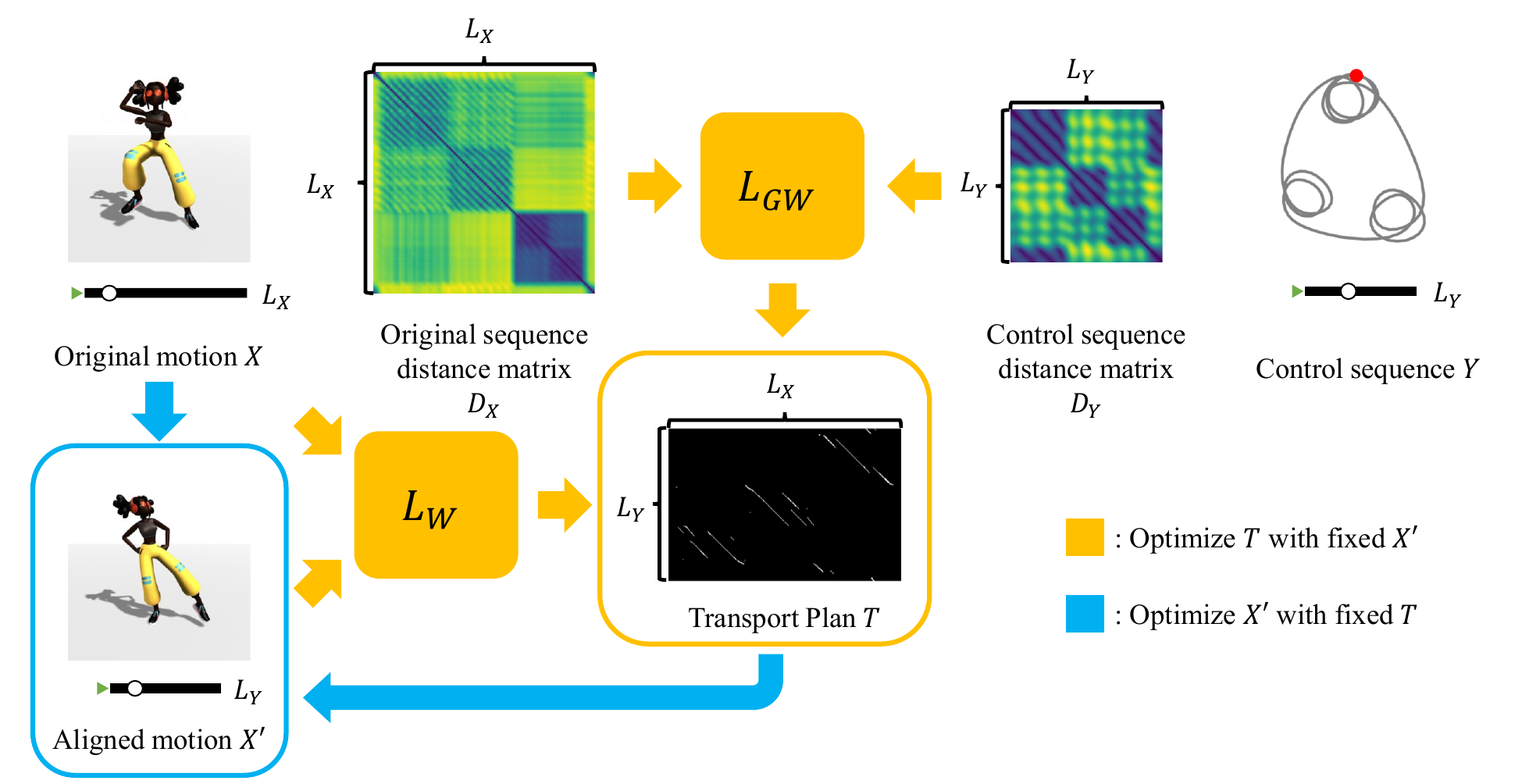}
    \caption{Explanation of the fused semi-unbalanced Gromov-Wasserstein (FSUGW) objective and algorithm to minimize it. $L_W$ constrains $X'$ to resemble $X$ via transport plan $T$ and $L_{GW}$ encourages $T$ to be metric-aligning, which leads to structural similarity between $X'$ and $Y$. We optimize FSUGW objective with alternating steps, where the first step optimizes $T$ with $X'$ fixed, and second step optimizes $X'$ over fixed $T$.}
    \label{fig:FSUGW}
\end{figure}

In the actual algorithm, we define the problem as the minimization of the fused semi-unbalanced Gromov-Wasserstein (FSUGW) loss $L_{FSUGW}$. Our formulation of $L_{FSUGW}$, as presented in Equations (1), (2), and (3), along with the projected mirror-descent algorithm employed as a component of our framework, follows the formulations in previous work~\cite{xu_temporally_2024}. By minimizing this objective via MAMM, $X'$ is optimized as described. Our contribution lies in integrating coarse-to-fine optimization (Sec.~\ref{subsubsec: method_coarse_to_fine}) and the patch extraction and blending process into the optimization loop to ensure temporal consistency (Sec.~\ref{subsubsec:FSUGWBlock}). Fig.~\ref{fig:FSUGW} provides an overview of the framework.

\subsubsection{Motion Patch Extraction}
To enable fine-grained optimization, we follow previous patch-based generation methods~\cite{li_example-based_2023, elnekave_generating_2022} by splitting the control sequence $Y$, original motion $X$, and aligned motion $X'$ into overlapping temporal patches with a stride of one frame. These patches are labeled as $\tilde{Y}$, $\tilde{X}$, and $\tilde{X'}$, respectively. Note that $\tilde{Y}$ and $\tilde{X}$ remain fixed, while $\tilde{X'}$ is updated at every step of the optimization loop ($\tilde{X'}\leftarrow$  ExtractPatches$(X')$ ). Measuring the FSUGW loss at the patch level ensures the capture of local temporal structures and ensures continuity.

Hereafter, we denote the length of the patch sequences as $L_X=|\tilde{X}|$ and $L_Y=|\tilde{Y}|=|\tilde{X'}|$.

\subsubsection{Wasserstein Loss}
\label{subsec:wd}

The Wasserstein loss focuses on establishing direct correspondence between the two sets, $X'$ and $X$. It is defined as follows:  
\[
L_{W}(\tilde{X'},\tilde{X},T) = \sum_{x'_i \in \tilde{X'}, x_k \in \tilde{X}} d_X(x_i', x_k) T_{i,k}. \tag{1}
\]

where $T \in \{T \in \mathbb{R} ^ {L_Y \times L_X}| T\geq0,  T\boldsymbol{1}=a,a=\boldsymbol{1}/L_Y\}$ denotes the transport plan matrix (coupling matrix) between $\tilde{X'}$ and $\tilde{X}$, and $d_X$ is a normalized distance function in the domain of $\tilde{X}$ and $\tilde{X'}$. We use the L2-norm as the distance function, normalizing it by either the maximum or mean distance, depending on the task. By minimizing $\min_{T}L_{W}$ w.r.t $X'$, we encourage each patch in $X'$ to move closer to its corresponding patch in $X$ via the coupling $T$, resulting in the aligned motion resembling the original motion.

\subsubsection{Gromov-Wasserstein Loss}

The Gromov-Wasserstein (GW) loss evaluates the alignment of internal structures between two datasets, specifically the control $Y$ and the original $X$. The loss, $L_{GW}(\tilde{Y}, \tilde{X}, T)$, is expressed as follows:
\[
L_{GW}(\tilde{Y}, \tilde{X}, T) = \sum_{\substack{y_i, y_j \in \tilde{Y} \\ x_k, x_l \in \tilde{X}}} |d_Y(y_i, y_j) - d_X(x_k, x_l)|^2 T_{i,k} T_{j,l}. \tag{2}
\]
where $d_Y$ is the normalized distance function in the domain of $Y$. It is important to note that we do not use the distance between $x \in \tilde{X}$ and $y \in \tilde{Y}$ at all.

Minimizing $L_{GW}$ encourages the transport plan $T$ to be metric-aligning, ultimately promoting structural alignment between $X'$ and $Y$ when combined with the Wasserstein loss described in Sec.~\ref{subsec:wd}.

\subsubsection{Fused Semi-Unbalanced Gromov-Wasserstein (FSUGW) Block}
\label{subsubsec:FSUGWBlock}
We combine the Wasserstein loss and GW loss to define and optimize $L_{FSUGW}$ as an instance of the entropic fused semi-unbalanced Gromov-Wasserstein problem~\cite{xu_temporally_2024}:
\[
\begin{aligned}
X'^*, T^* = \underset{X', T}{\operatorname{argmin}} \quad & L_{\text{FSUGW}} (\tilde{X'}, \tilde{X}, \tilde{Y}, T) \\
\text{where} \quad L_{\text{FSUGW}} (\tilde{X'}, \tilde{X}, \tilde{Y}, T) =\ & \alpha \cdot L_{\text{GW}}(\tilde{Y}, \tilde{X}, T) \\
 +\, &(1 - \alpha) \cdot L_{\text{W}}(\tilde{X'}, \tilde{X}, T) \\
 +\, &\lambda \cdot D_{\text{KL}}(T^{\top} \boldsymbol{1} \,\|\, \boldsymbol{b}) -\, \epsilon \cdot H(T), \\
\text{subject to} \quad & T \boldsymbol{1} = \boldsymbol{a}, \quad T \geq 0. \\
\end{aligned} \tag{3}
\]

Here, we define
$a = \boldsymbol{1}_{L_Y} / {L_Y}, \quad b = \boldsymbol{1}_{L_X} / {L_X}$,
where $L_Y$ and $L_X$ are the number of data (frames) in $\tilde{Y}$ (or $\tilde{X'}$) and $\tilde{X}$, respectively. The term $H(T)=-\sum_{i,j}T_{i,j}log(T_{i,j})$ represents the soft entropy constraint, which facilitates optimization via the projected mirror-descent algorithm~\cite{solomon_entropic_2016, xu_temporally_2024}. The soft marginal constraint, defined as the $D_{KL}$ term, encourages bi-directional fidelity of the distribution of the aligned motion $X'$ to that of $X$. While soft constraints are imposed on one side, we enforce hard constraints on the other side of the marginals, $T\boldsymbol{1} = a$,
ensuring that $T$ satisfies the specified marginals. Intuitively, this constraint ensures that every part of the aligned motion $X'$ corresponds to some part of the original motion $X$. The effects of the parameters $\alpha$ and $\lambda$ will be discussed in Sec.~\ref{subsec: discussion_parameter_effects}.

Optimizing $L_{FSUGW}$ w.r.t $T$ and $X'$ is similar to the (fused unbalanced) Gromov-Wasserstein Barycenter problem~\cite{peyre_gromov-wasserstein_2016, thual_aligning_2022}, which is typically solved using a block coordinate descent algorithm. Inspired by this, our proposed FSUGW block optimizes $L_{FSUGW}$ by alternating the following steps:  
\begin{enumerate}
    \item \textbf{Extract patches:} Update the patches $\tilde{X'}$ using the current value of $X'$,
    \[
    \tilde{X'} \gets \text{ExtractPatches}(X')
    \]
    \item \textbf{Optimize $T$ with fixed $X'$:} Given $X'$, we find $T$ as a local minimum of $L_{FSUGW}$ by applying the projected mirror-descent algorithm, identical to the version in~\cite{xu_temporally_2024} except for the parameters. At each stage, $T$ is initialized as $\mathbf{a}\mathbf{b}^T$, following previous practices.
    \item \textbf{Optimize $X'$ with fixed $T$:} With $T$ fixed, update $X'$ by matching weighted by the transport plan $T$ and blending the overlapping regions through averaging.
    \[
    X' \gets \text{BlendPatches}(T \cdot \tilde{X} \cdot L_Y).
    \]
\end{enumerate}
The patch extraction and blending process in the optimization loop ensures that spatio-temporally adjacent patches in $X$ are encouraged to remain adjacent in $X'$. If $X'$ is formed by blending spatio-temporally distant patches, it may result in highly unnatural poses in the overlapping regions of those patches. However, such cases are naturally excluded during optimization, as they increase $L_W$, which the algorithm minimizes.

In practice, this algorithm typically converges after several iterations to produce visually appealing results across various applications, as demonstrated in Sec.~\ref{sec:applications}. For these steps, we set $M=20$ as the number of iterations.

\subsubsection{Coarse-to-Fine Motion Alignment}
\label{subsubsec: method_coarse_to_fine}
Owing to the non-convex nature of GW problems~\cite{solomon_entropic_2016}, proper initialization of the FSUGW block is crucial for guiding the optimization process toward high-quality aligned motion.
To ensure effective initialization, we adopt a coarse-to-fine optimization strategy, as used in previous works~\cite{li_ganimator_2022, li_example-based_2023, elnekave_generating_2022}. This approach begins at a coarse level, with the initial transport plan $T$ computed by minimizing only the $L_{GW}$ term, and progressively refines the motion through upsampling until reaching the final stage:
\begin{enumerate}
    \item \textbf{Upsampling}: Upsample the output $X'^{k-1}$ from the previous stage to serve as the initial guess for stage $k$.
    \item \textbf{FSUGW Optimization}: Apply the FSUGW block using $Y^k, X^k$ to refine $X'^k$.
\end{enumerate}

After the final stage $k = N$, $X'^N$ achieves a high-quality alignment with the control sequence $Y$, while preserving the content of the original motion $X$. In our examples, the coarsest level is set to be $4$ times shorter than the final length, and $N=6$ stages are used to progressively upsample the sequence. The pseudocode for the entire MAMM algorithm is provided in Alg.~\ref{algo:all}.

\begin{algorithm}[t]
\caption{MAMM (Metric-aligning motion matching)}
\label{algo:all}
\KwIn{control sequence $Y$, original motion $X$, parameters $\alpha$, $\lambda$, $\epsilon$, number of stages $N$, number of iterations $M$}
\KwOut{aligned motion $X'$}

// Coarse-to-fine Motion Alignement

\For{$k \gets 0$ \KwTo $N$}{
    $X' \gets (k=0)\,?\,\text{Initialize()}:\,\text{Upsample}(X')$\;
    
    $X^k, Y^k \gets$ Downsample($X, Y, k$)\;

    // FSUGW Block
    
    \For{$m \gets 0$ \KwTo $M$}{
        $(\tilde{X'}, \tilde{Y}, \tilde{X}) \gets$ ExtractPatches($X', Y^k, X^k$)\;
        
        $a \gets \boldsymbol{1}/{L_Y},$ \quad 
        $b \gets \boldsymbol{1}/{L_X},$ \quad $T \gets (m=0)\,?\,\mathbf{ab}^T:\,T$\;
        
        // Optimize $T$ via projected mirror descent
        
        $T \gets$ SolveFSUGW($\tilde{X'}, \tilde{X}, \tilde{Y}, T, a, b, \alpha, \lambda, \epsilon$)\; 
        
        // Optimize $X'$ via matching
        
        $X' \gets$ BlendPatches($T \cdot \tilde{X} \cdot L_Y$)\;
    }
}
\Return $X'$\;
\end{algorithm}
\subsection{Handling Additional Space-Time Constraints}

\subsubsection{Soft Keyframes by Example Pairs}
We introduce the concept of “soft keyframes” by adding \textit{example patch pairs} into the FSUGW optimization. These example patch pairs represent sample-to-sample correspondences between the domains of $\tilde{Y}$ and $\tilde{X}$. The patch samples do not necessarily have to be part of the original or control sequence; the only requirement is that they belong to the same domain. We incorporate these example patch pairs into the optimization by applying the following substitution:
\begin{align*}
    \tilde{X} \gets \bigl[\tilde{X},\, X_{\mathrm{example}}\bigr], 
    \tilde{X}' \gets \bigl[\tilde{X}',&\, X_{\mathrm{example}}\bigr], 
    \tilde{Y} \gets \bigl[\tilde{Y},\, Y_{\mathrm{example}}\bigr], \\
    a \gets \bigl[a,\, \tfrac{w_{\mathrm{example}} \cdot \mathbf{1}}{L_{\mathrm{example}}}\bigr],&
    b \gets \bigl[b,\, \tfrac{w_{\mathrm{example}} \cdot \mathbf{1}}{L_{\mathrm{example}}}\bigr],
\end{align*}
where $X_{\mathrm{example}}$ and $Y_{\mathrm{example}}$ denote the example patch pairs, $w_{\mathrm{example}}$ denotes the weight of the soft keyframe constraints, and $L_{\mathrm{example}}$ denotes the number of example patches. Additionally, we constrain the transport matrix $T$ such that all masses assigned to the example patches in $\tilde{X}$ (or $\tilde{X}'$) are transported exclusively to their corresponding example patches, and vice versa. This ensures that these soft keyframes do not interfere with the overall distribution derived from the original motion $X$, which is reflected in the aligned motion $X'$ through the soft marginal constraint $D_{\mathrm{KL}}\bigl(T^\top\mathbf{1}\,\|\,\mathbf{b}\bigr)$.
The difference between these soft keyframes and traditional (hard) keyframes, along with illustrative examples, is discussed in Sec.~\ref{sec:discussions}.
\subsubsection{Other Constraints}
Because a portion of our method relies on patch extraction and blending, it is compatible with various space-time control methods. In our implementation, we integrate \textit{hard keyframing} and \textit{infinite loop} constraints, as outlined in GenMM~\cite{li_example-based_2023}.
\paragraph{Hard Keyframes.}
Hard keyframes are specified pairs of patches in $X$ and $Y$ that must be coupled. When imposing a hard keyframe constraint, we fix the specified segments of both $X'$ and $T$ at every step of the FSUGW optimization process, including during the projected mirror-descent routine. This ensures that the predetermined keyframes remain unchanged throughout the alignment process.
\paragraph{Infinite Looping.}
When imposing an infinite loop constraint, the last patch of the aligned motion sequence is enforced to match the first patch of the same sequence, ensuring a seamless loop. Following the practice in GenMM~\cite{li_example-based_2023}, we achieve this by blending the overlapping region between the end and start patches consistently.

Alignment results with these constraints can be observed in the supplemental video along with the demonstration of our sketch-to-motion applications.

\begin{figure*}[t]
    \centering
    \includegraphics[width=\linewidth, trim=0 0 0 0, clip]{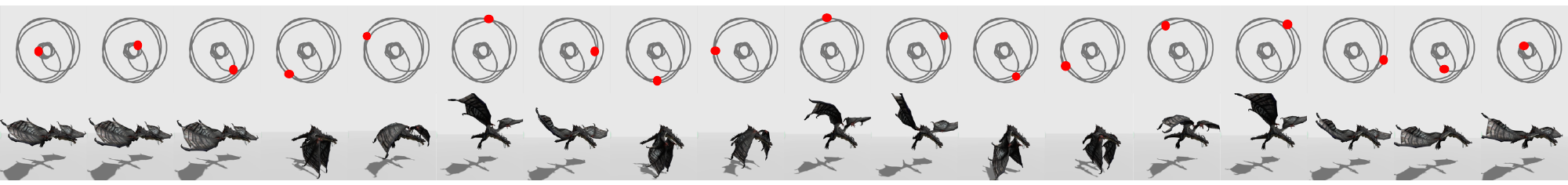}
    \caption{Demonstration of character motion control using sketched curve. Character's movement follows abstract structure of the curve. Additional examples are available in supplementary video.}
    \label{fig:sketch}
\end{figure*}
\section{Applications}
\label{sec:applications}
In this section, we demonstrate how the MAMM framework can be applied to various input control sequences. All use cases are realized by splitting each control sequence into patches, computing the distance matrix among the patches, and adjusting the optimization parameters; the underlying algorithm itself remains unchanged. All data used in the examples are sourced from Mixamo~\cite{noauthor_mixamo_nodate} or Truebones~\cite{noauthor_truebones_nodate}, which contain various motions and skeletons of humanoids and non-humanoids, respectively. The sequence lengths in our experiments range from approximately  50 to 600 frames at 30 fps. The patch size of the original motion was set to 11 frames, with a stride of one frame. Patch extraction from the control sequence used the same temporal duration and interval. Additionally, we used the AIST dance database~\cite{aist-dance-db} for music data and the BEAT dataset~\cite{liu2022beat} for speech data.

We implemented our applications on a personal computer equipped with an NVIDIA GeForce RTX 4080 GPU. Each generation typically finishes within 10 s but may take longer depending on the sequence length. The parameter values used in our applications are detailed in the supplemental materials. We also include a supplemental video to facilitate the visual evaluation of the alignment results.
\subsection{Sketch-to-Motion (2D Curve-to-Motion)}
In this application, the control sequence is a temporal series of two-dimensional data points provided by the user. To enhance the user experience, we have developed a dedicated user interface (UI) for sketch-based motion control. In this UI, the user draws a 2D curve on the left-hand canvas and observes the aligned animation on the right. The user can iteratively refine the animation by redrawing the curve, placing soft or hard keyframes on the canvas, or enabling infinite looping. The 2D points along the curve are resampled at constant time intervals as patches and then fed into the alignment algorithm. To reduce the impact of unintended noise in the input, we apply Gaussian smoothing before processing the curve in our framework. 

The user does not need prior knowledge of which part of the canvas corresponds to specific frames of the animation. As such, the exact location of the curve does not matter. Instead, the system analyzes the overall shape and movement of the curve and maps it to the general structure of the motion. For example, a large circle is automatically matched with a large cyclic motion, while a small circle corresponds to a smaller motion. Note that we compute a mapping from a sampled point on the curve to a motion patch. This is not a continuous mapping from an arbitrary spatial position to a pose. Therefore, even if the curve passes through the same position (i.e., a crossing), the pose at that location on the curve can differ from the pose at the same spatial location at a different point in time. This implicit matching is effective in scenarios where exact poses are less critical and dynamics are more important. If precise mapping is needed, the user can specify it by adding soft or hard constraints. An example result is shown in Fig.~\ref{fig:sketch} and Fig.~\ref{fig:sketch-keyframe}. Additionally, we performed a user study to gather feedback on this tool, which is discussed in Sec.~\ref{sec:userfeedback}.
\subsection{1D Waveform-to-Motion}
\begin{figure*}
    \centering
    \includegraphics[width=\linewidth]{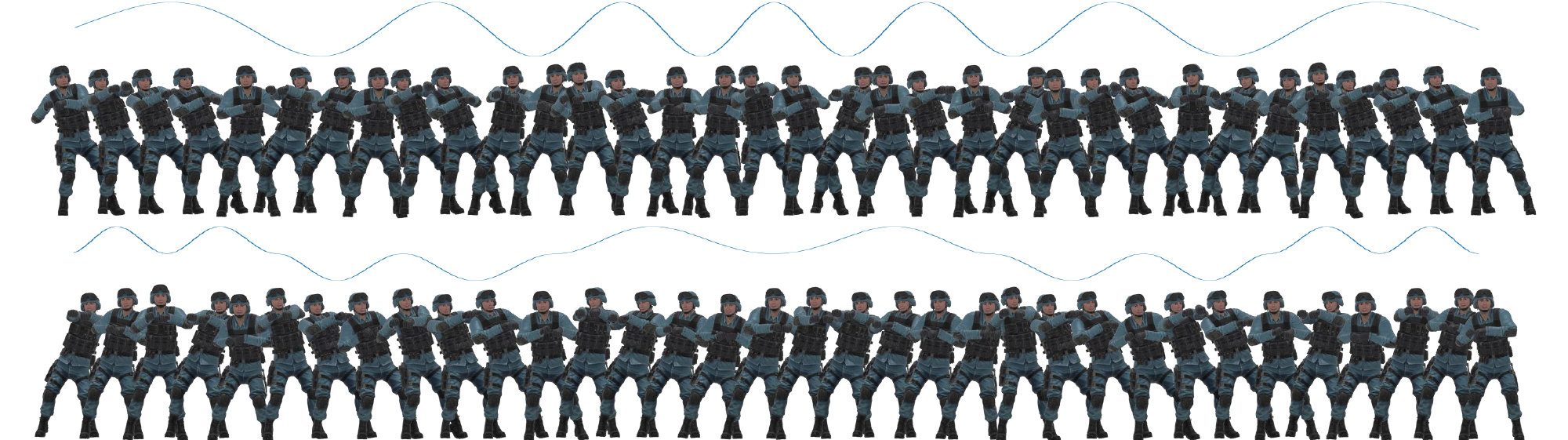}
    \caption{Examples of controlling motion by waveforms. We used (1) frequency-varying sine wave and (2) center-alternating sine wave to control the dance motion's frequency and phase. For more examples, please refer to the attached video.}
    \label{fig:waveform_to_motion}
\end{figure*}

Here, the control sequence is a temporal series of one-dimensional scalar values. We demonstrate how periodic motion patterns can be controlled using time-varying periodic waveforms. A low-frequency sine wave generates slow periodic motion, while a high-frequency sine wave produces fast periodic motion. Sharp corners in the waveform result in sharp turns in the motion. Additionally, we use a sine wave with an alternating center as input to achieve complex periodicity control. The results are provided in Fig.~\ref{fig:waveform_to_motion} and the supplemental videos.
\subsection{Motion-by-Numbers}
Drawing inspiration from successful image synthesis pipelines guided by segmentation labels~\cite{isola_image--image_2017}, we align motion to a user-specified temporal sequence of segmentation labels, each represented as a one-hot class vector. With MAMM, the original motion sequence does not require any precomputed  or annotated segmentation labels; instead, the algorithm automatically aligns each segment according to the input labels. The user does not need prior knowledge of which label corresponds to which motion. However, when necessary, the user can explicitly specify the mapping by providing soft or hard constraints. Results are shown in the center row of Fig.~\ref{fig:alpha_comparision} and Fig.~\ref{fig:lambda_comparision}, as well as in the supplemental video.
\subsection{Motion Control by Audio}
\begin{figure*}
    \centering
    \includegraphics[width=\linewidth]{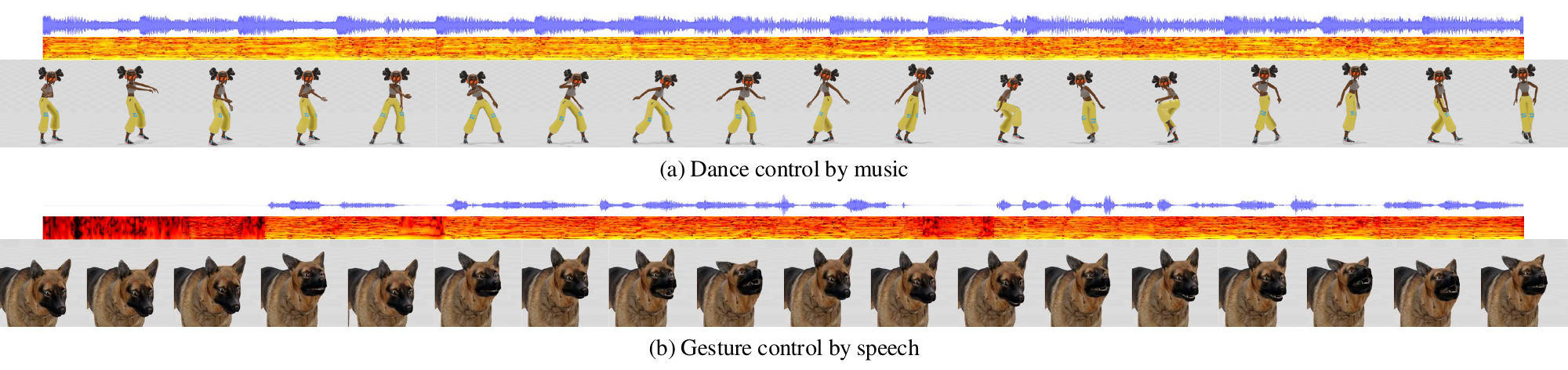}
    \caption{Examples of motion controlled by audio data. We tested two scenarios: (a) dance movements controlled by music and (b) gestures (barking) controlled by speech. Here audio data is represented in both waveforms and MFCCs, but we only used MFCCs as input. Synchronization between motion and intensity, beat, or style of audio was observed. For more details, please refer to supplemental video.}
    \label{fig:audio}
\end{figure*}

As mentioned in Sec.~\ref{subsec:rel_dance}, audio-driven motion control has been extensively studied in the motion generation community. We focus on a scenario where the available example sequences are relatively short (3–20 s  in duration) and lack paired audio or auxiliary annotations. The audio input is represented as 40-dimensional MFCC features, extracted using the Librosa library~\cite{mcfee_librosalibrosa_2024}. Both speech and music data are used as audio controls, while short segments of gesture or dance motion serve as the original motion. As presented in Fig.~\ref{fig:audio}, our method naturally synchronizes the resulting motion with style, beat, or volume changes in the audio, even when the original motion style differs from that in the audio. Notably, this approach does not require any training or specialized, hand-tuned optimization.
\subsection{Motion-to-Motion Alignment}
\begin{figure*}
    \centering
    \includegraphics[width=\linewidth]{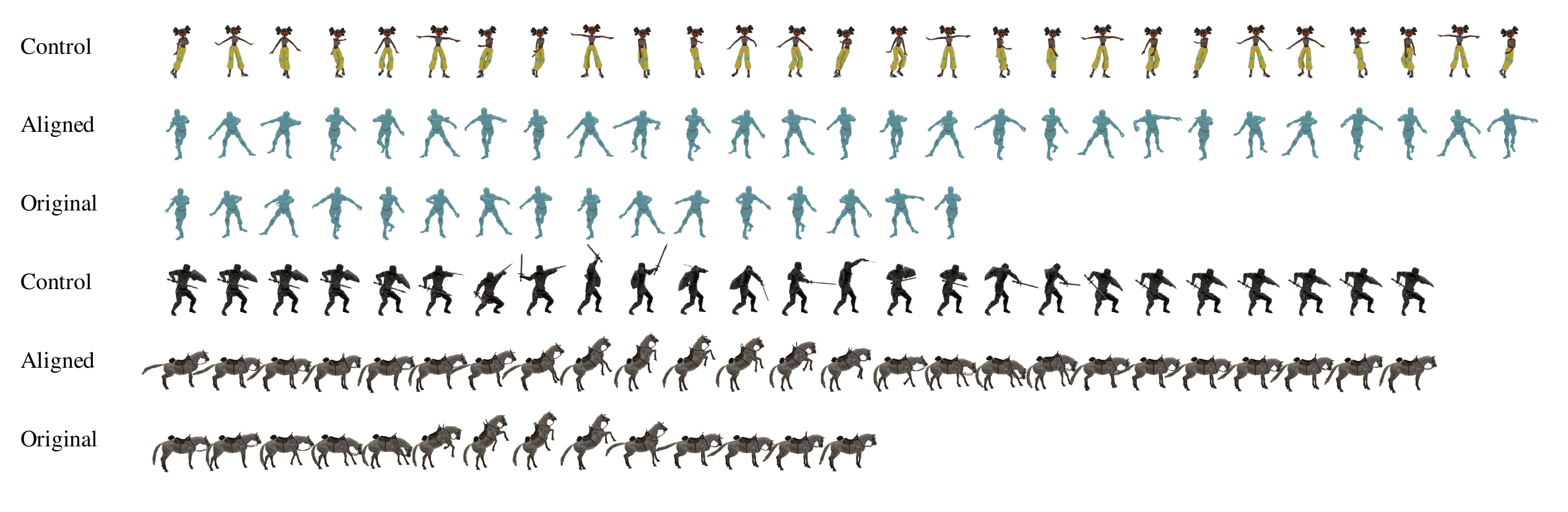}
    \caption{Examples of motion-to-motion alignment, where original motion is synchronized with control motion. First example aligns stepping motion to step with different skeletal structure and style, matching both frequency and phase. Second example aligns nonperiodic combat sequence involving human and horse, demonstrating synchronized timing of combat actions.}
    \label{fig:retarget}
\end{figure*}

We applied MAMM to cross-skeletal motion-to-motion alignment, where one motion sequence is matched to another with a different skeletal structure~\cite{li_walkthedog_2024}. An example is shown in Fig.~\ref{fig:retarget}. Leveraging its metric-alignment principle, our method effectively handles both periodic motions, such as walking, and non-periodic activities, such as combat moves, without requiring additional training or customized feature engineering.

\section{Discussions}
\label{sec:discussions}
\subsection{Parameter Effects}
\label{subsec: discussion_parameter_effects}

\begin{figure}
    \centering
    \includegraphics[width=0.96\linewidth]{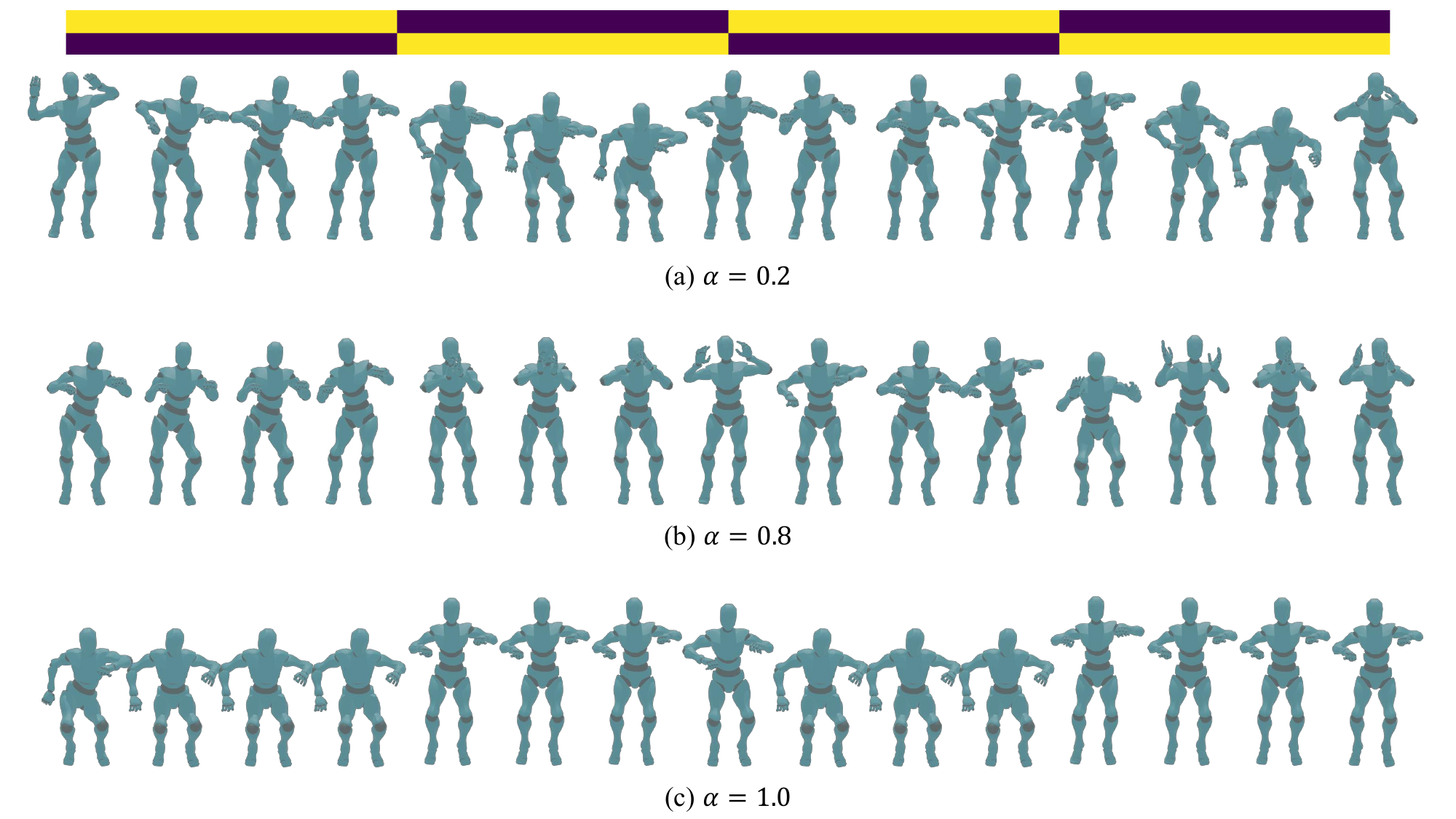}
    \caption{Aligned motions for various $\alpha$ values with $\lambda$ and $\epsilon$ fixed to $0.05$ and $1.0$, respectively. Small $\alpha$ (e.g., 0.2) results in natural motion but occasionally ignores control inputs, such as style changes within the same segmentation label. Conversely, $\alpha$ value of 1.0 leads to aligned motions with segments that exhibit no movement.}
    \label{fig:alpha_comparision}
\end{figure}

\begin{figure}
    \centering
    \includegraphics[width=0.96\linewidth]{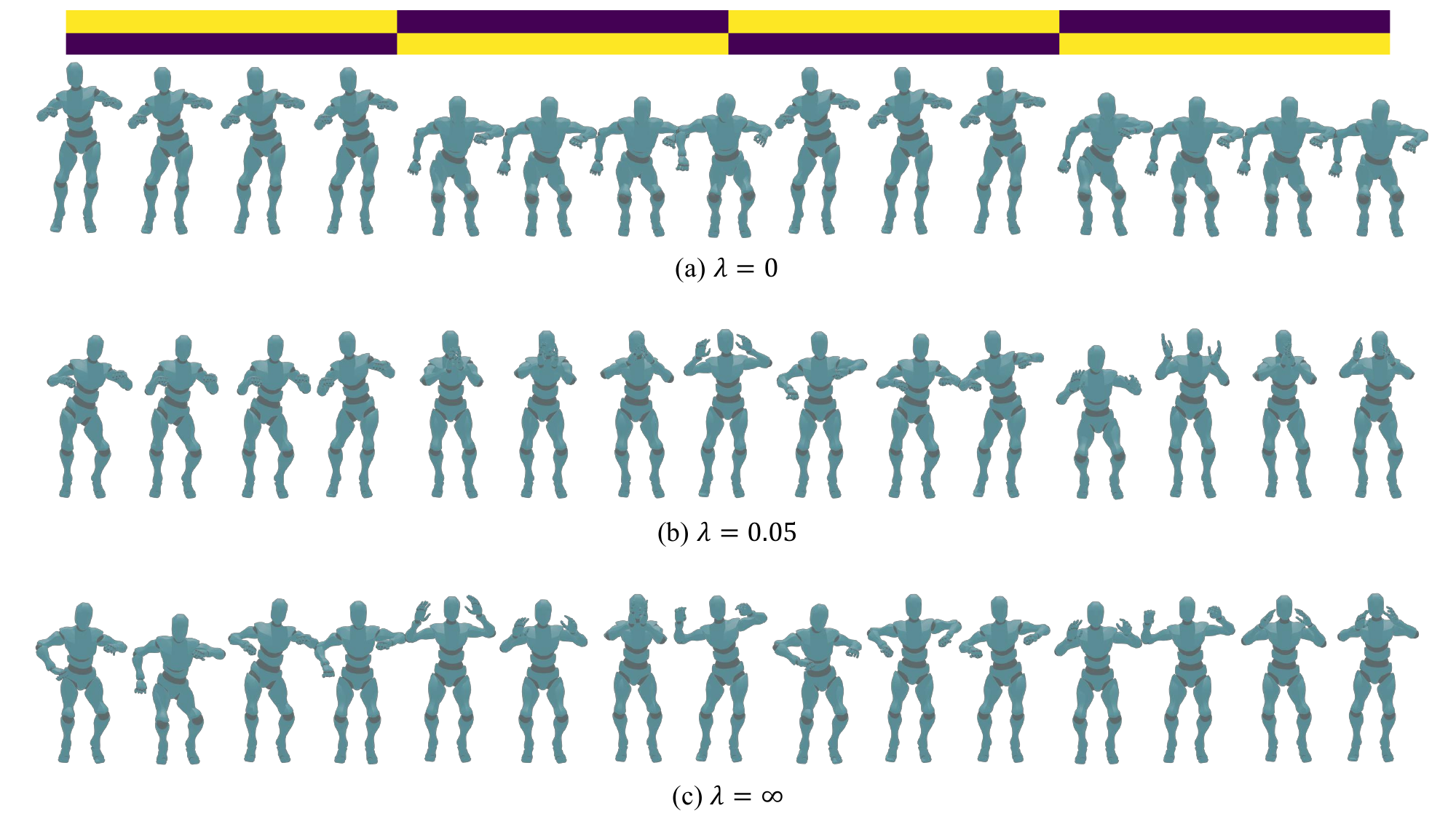}
    \caption{Aligned motions for various $\lambda$ values with $\alpha$ fixed at $0.8$. When $\lambda$ is small (e.g., 0), aligned motion includes segments where poses remain unchanged, deviating from the overall dynamics of the original motion. By contrast, large $\lambda$ (e.g., $\infty$) ensures that distribution of aligned motion patches closely matches original motion, often at the expense of adhering to control sequence.}
    \label{fig:lambda_comparision}
\end{figure}

We explain the effect of the main parameters in our framework ($\alpha$ and $\lambda$). See Fig.~\ref{fig:alpha_comparision}, Fig.~\ref{fig:lambda_comparision} or supplemental materials to see examples.

The parameter $\alpha$ balances the Wasserstein loss $L_W$ and the Gromov-Wasserstein loss $L_{GW}$. The $L_{GW}$ term maintains structural similarity between the control sequence and the aligned motion, while $L_W$ enforces local, patchwise fidelity to the original motion. Increasing $\alpha$ amplifies the influence of the control input but may lead to unnatural outputs if set too high. 

The parameter $\lambda$ scales the soft marginal regularization term $D_{\mathrm{KL}}\bigl(T^\top \mathbf{1}\,\|\,\mathbf{b}\bigr)$. Intuitively, a larger $\lambda$ forces the aligned motion to more closely maintain the overall distribution of the original motion, potentially diminishing the influence of the control input if $\lambda$ is set too high.

We recommend selecting relatively large values for $\alpha$ and small values for $\lambda$, such as $\alpha = 0.8$ and $\lambda = 0.05$. After evaluating the initial results, these parameters can be fine-tuned by decreasing $\alpha$ or increasing $\lambda$ to enhance naturalness, or vice versa, depending on the desired outcome. Additionally, we have observed that setting $\epsilon = 1$ works well for most settings; however, adjustments may be needed based on the size or complexity of the data.

\subsection{Coarse-to-Fine Motion Alignment}
While adopting a coarse-to-fine optimization approach generally improves results, it becomes especially important when the distance matrix of the control sequence is more complex, such as in motion-to-motion alignment scenarios. In the supplementary video, we highlight failure cases that arise when using a one-stage (non-coarse-to-fine) optimization method.
\subsection{Soft Keyframes and Hard Keyframes}
As discussed in Sec.~\ref{sec:method}, our MAMM framework accommodates both \textit{soft keyframing} and \textit{hard keyframing}. In this section, we use the sketch-to-motion application to illustrate these keyframe concepts
in a practical context.

Soft keyframes are defined as pairs of patch samples drawn from the control domain and the original motion domain. By contrast, hard keyframes pair a specific point in time (on the temporal axis) with a patch in the original motion domain. Consequently, in the \textit{sketch-to-motion} setting, soft keyframes can be placed anywhere on the canvas, whereas hard keyframes must be associated with a specific time frame and positioned directly on the curve.

\begin{figure*}[h]
    \centering
    \includegraphics[width=0.8\linewidth, trim=0 50mm 0 0, clip]{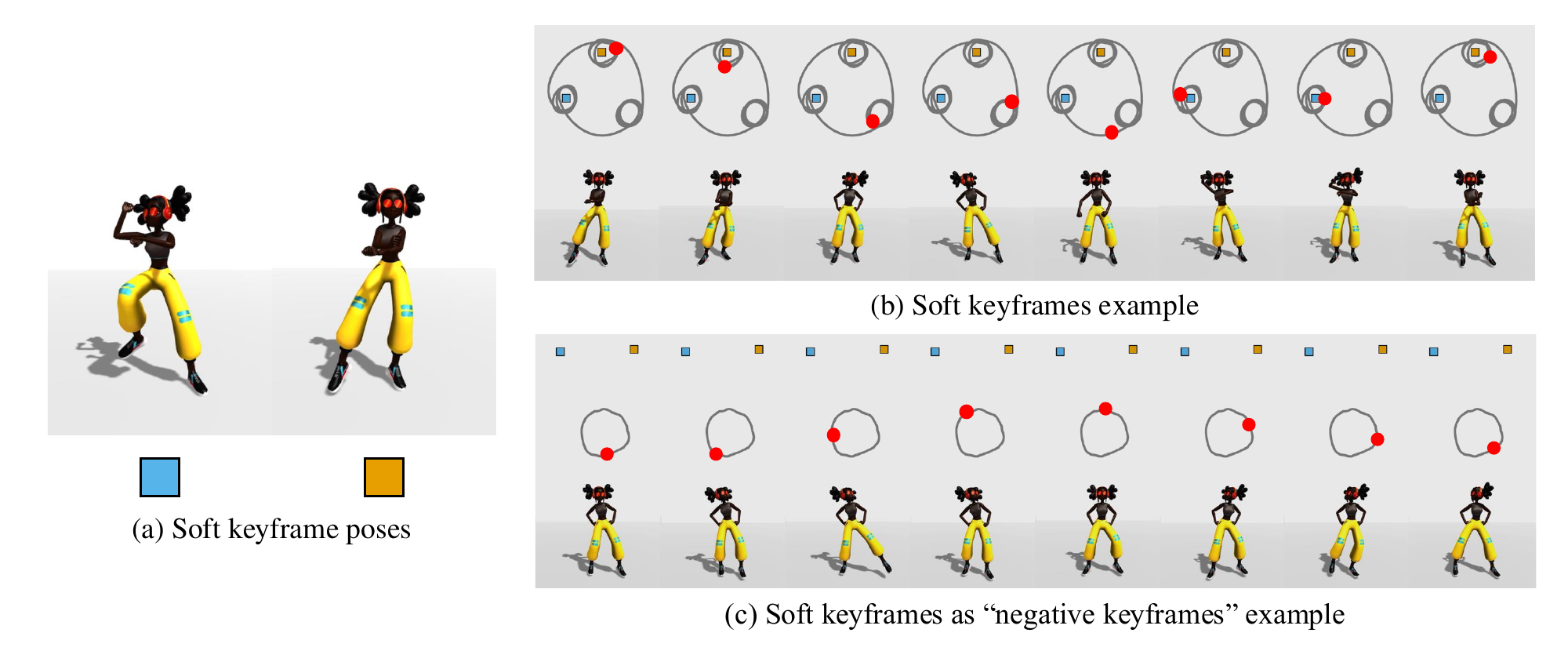}
    \caption{Example of controlling aligned motion with soft keyframes. (a) Users can specify keyframe poses, such as "hands-up horse rider" and "hands-down horse rider," using our interface. For simplicity, users select poses from original sequence, although our method does not impose strict constraints on keyframe selection. (b) When motion curve approaches keyframe, algorithm ensures that corresponding poses in aligned motion closely match keyframe poses. (c) Conversely, when curve is distant from keyframe, algorithm selects alternative poses, such as "side-step" poses, which differ from keyframe poses. This illustrates how soft keyframes can influence motion in both positive and negative contexts.}
    \label{fig:sketch-keyframe}
\end{figure*}

This distinction enables an interesting application of soft keyframes: they can act as “negative keyframes”. As illustrated in Fig.~\ref{fig:sketch-keyframe}, if the user’s drawn curve is positioned far from keyframes representing the “hands-up horse rider” and “hands-down horse rider” poses, the algorithm instead generates a looped “side step” sequence that stays distant from both keyframe poses.

\section{User Feedback}
\label{sec:userfeedback}

\begin{figure*}
    \centering
    \includegraphics[width=\linewidth]{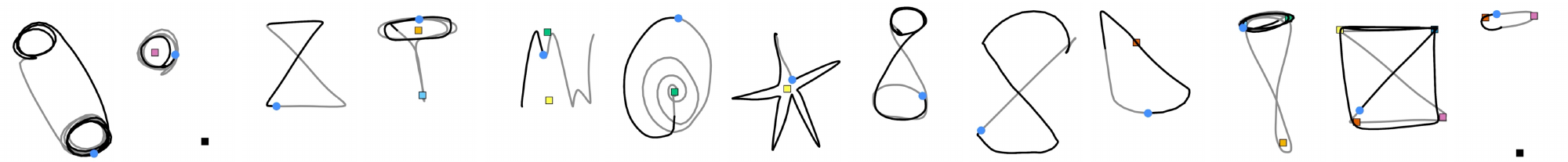}
    \caption{Examples of sketches created by user study participants. Colored squares represent keyframes. For the animated results of each curve, please refer to the supplementary materials.}
    \label{fig:user-sketches}
\end{figure*}

We invited five participants to test the sketch-to-motion application to obtain feedback. The participants were computer science students with no prior experience in character animation editing. During the user study session, we provided a 10-minute tutorial followed by 15 minutes of guided practice. Examples of sketches created by participants are shown in Fig.~\ref{fig:user-sketches}. Finally, the participants were tasked with arranging an animation to match a given video. All participants successfully recreated a motion sequence similar to that shown in the video.

We generally received positive feedback, such as ``The application can be used even without experience in animation editing.'' and ``I found changing animations with keyframes to be very convenient.'' However, some participants noted difficulties, stating ``I found the relationship between sketches and animations to be a bit unclear.'' They also suggested improvements, including ``Suggesting keyframe candidates could reduce editing time,'' and ``A feature to completely remove unwanted keyframes (hard negative keyframes) would be useful.'' The animations created by the participants and detailed survey results are included in the supplementary materials.

\section{Limitations and Future Works} 


\paragraph{Real-time control.} 
In our formulation, the user needed to provide a complete control sequence before computing the alignment, making real-time interaction impossible. Future work could focus on developing fast approximation techniques for MAMM to enable real-time user control.

\paragraph{Ambiguity to Rigid Transformation and Symmetry}
Since the $L_{GW}$ term relied on a metric, it was  invariant to rigid transformations and symmetries in the patch space. For example, the method could not distinguish between stepping to the right and stepping to the left in symmetrical step motions, which could result in outputs differing from users' expectations. However, tools such as keyframes were provided to address and control this issue.

\paragraph{Pairwise Distance Scaling.} 
A key challenge lay in optimally designing and normalizing the distance functions, denoted by $d_X$ and $d_Y$. Although these distance functions were normalized using the max or mean value in our experiments, we also attempted to learn a scalar scaling factor \( s \) for $d_{\tilde{Y}}$ within the optimization loop by setting its value to the global minimum of $L_{GW}$ with $T$ fixed.
However, this approach did not yield consistent improvements. Future work should investigate more robust matrix-scaling strategies, particularly for larger datasets that exhibit multiple clusters.
\paragraph{Scalability Concerns.}
MAMM’s computational complexity increased both spatially and temporally with larger input sequences. The largest example of $T$ we tested involved motion with $594$ frames ($19.8$ s) and control audio with $1441$ MFCC features ($48.04$ s), which took $7$ s to compute. However, as modern motion datasets sometimes contain  over $100$ thousands of frames, storing and operating on such high-dimensional distance matrices could become prohibitive. Developing more scalable approaches---such as improved hierarchical techniques or approximations to FSUGW---constitutes an important direction for future research.
    


\begin{acks}
We thank the anonymous reviewers for their insightful feedback.
We would like to thank Takaaki Shiratori and Peizhuo Li for the valuable feedback and helpful discussions on this project.
This work was supported by Japan Science and Technology Agency (JST) as part of Adopting Sustainable Partnerships for Innovative Research Ecosystem (ASPIRE), Grant Number JPMJAP2401.
\end{acks}

\bibliographystyle{ACM-Reference-Format}
\bibliography{sample-bibliography}

\appendix


\end{document}